\documentclass[proceedings, preprint]{rmaa}

\usepackage{paralist}

\usepackage{psfrag,color}
\usepackage{times}
\usepackage{graphicx}
\usepackage{ulem}
\usepackage{epsfig}
\usepackage{natbib}

\newcommand{\msun}{\mbox{$M_\odot$}}
\newcommand{\rsun}{\mbox{$R_\odot$}}

\def\be{\begin{eqnarray}}
\def\ee{\end{eqnarray}}
\def\bi{\begin{itemize}}
\def\ei{\end{itemize}}
\def\lsim{\mathrel{\rlap{\lower3pt\hbox{\hskip1pt$\sim$}}
     \raise1pt\hbox{$<$}}} 
\def\gsim{\mathrel{\rlap{\lower3pt\hbox{\hskip1pt$\sim$}}
     \raise1pt\hbox{$>$}}} 

\SetYear{2013}
\SetConfTitle{Magnetic Fields in the Universe 4}

\title{GRBs from Weakly-Magnetized, Slowly-Rotating Stars in Binaries} 

\author{Enrique Moreno M\'endez,\altaffilmark{1}}

\altaffiltext{1}{Instituto de Astronom\'ia, Universidad Nacional Aut\'onoma de M\'exico, Circuito Exterior, Ciudad Universitaria, Apartado Postal 70-543, 04510, Distrito Federal, M\'exico. (enriquemm@astro.unam.mx).}

\shortauthor{Moreno M\'endez}
\shorttitle{RevMexAA(SC) Demo Document}

\listofauthors{Enrique Moreno M\'endez}
\indexauthor{Enrique Moreno M\'endez}

\abstract{The spin of a number of black holes (BHs) in X-ray binaries (XBs) has been predicted (and, in at least three cases, confirmed by observations) by using a binary stellar evolution model with Case-C mass transfer .  
The rotational energy of such BHs is sufficient to power up (long) gamma-ray bursts and hypernovae (GRBs/HNe) and still leave a Kerr BH behind.  
However, strong magnetic fields (B fields) and/or dynamo effects in the interior of a BH-progenitor star may be capable of rapidly depleting the angular momentum from the stellar core, hence, preventing the formation of a collapsar.  Thus, even if binaries can produce Kerr BHs, most of their rotation is acquired from accreting the stellar mantle, with a long delay between the formation of the BH and its spin up.  
Hence, not being good sources of GRBs.

We study the necessary conditions to produce GRBs by the progenitors of such BHs.  
Tidal-synchronization and Alfv\'en timescales are compared for B fields of different intensities threading trough He stars.  
We search  for a B-field range which allows tidal spin up all the way into stellar core but prevents its slow down during differential rotation phases. Energetics for producing a strong B field during core collapse, which allows for GRB central engines, are also estimated.
An observationally-reasonable choice of parameters is found.  
}

\resumen{Utilizando evoluci\'on estelar de sistemas binarios (con transferencia de masa Caso C) el esp\'in de varios agujeros negros (ANs) en binarias de rayos X ha sido predicho, observaciones posteriores han confirmado al menos tres de estas predicciones.  
La energ\'ia rotacional de \'estos ANs es suficiente para producir destellos de rayos gama (GRBs) largos acompa\~nados por explosiones de hipernova (HN) y a\'un as\'i dejar un AN de Kerr.
No obstante, campos magn\'eticos intensos y/o d\'inamos en el interior de la estrella progenitora del AN pueden extraer toda la reserva de momento angular en el n\'ucleo estelar previniendo la formaci\'on de un col\'apsar.
Esto es, a\'un en binarias capces de formar ANs de Kerr, la mayor parte de la rotaci\'on del AN puede venir del manto estelar, lo cual produce un largo retraso entre la formaci\'on del AN y la adquisici\'on de su esp\'in.
Este escenario descartar\'ia a estas binarias como posibles precursoras de GRBs.

En este trabajo se realiza un estudio sobre las condiciones necesarias para que las progenitores de ANs de Kerr en binarias de rayos X puedan producir GRBs.
Escalas de tiempo de sincronizaci\'on por fuerzas de marea y de Alfv\'en son comparadas para campos magn\'eticos de diferentes intensidades en estrellas de He.
Se esfect\'ua una b\'usqueda para un rango de campos magn\'eticos que permitan sincronizaci\'on por fuerzas de marea hasta el n\'ucleo estelar pero que prevengan la fuga de momento angular durante las subsiguientes etapas de rotaci\'on diferencial.  
Se estima tambi\'en el costo energ\'etico de la producci\'on de un campo magn\'etico durante el colapso gravitacional que permita la formaci\'on de un GRB.
Finalmente, se encuentra un rango de par\'ametros congruente con observaciones.
}

\addkeyword{binaries: general}
\addkeyword{black hole physics}
\addkeyword{gamma-ray bursts: general}
\addkeyword{stars: evolution}
\addkeyword{stars: magnetic field}
\addkeyword{stars: massive}

\begin{document}
\maketitle

\section{Introduction}
\label{sec:Intro}

Massive stars, those capable of leaving a massive core which may collapse into a black hole (BH) tend to lose a substantial amount of mass through winds.
The material leaving the star is at the surface, where most of the angular momentum of the star resides.
Thus, wind mass loss implies substantial spin angular momentum loss to these stars.
If the magnetic field is considerable at the stellar surface and beyond, the lever arm increases further helping
the spin down.
Furthermore, evolution into the giant and supergiant branch drasdtically increases the radii and the wind mass loss.
Hence, evolved stars lose most of the angular momentum left in the star after the main sequence.
This picture is enhanced by metallicity, as wind-mass loss is increases along with it.

Long gamma-ray bursts (LGRBs) are thought to be the product of a massive, rapidly-rotating, gravitationally-core-collapsing (GCC) star.
During GCC said stars collapse first from the poles as no centrifugal force prevents the material from falling.
This clears the way for jets to propagate along the rotational axis of the star as a lower density region, and lower resistance path is created.
It is also likely that during GCC differential rotation ensues and any fossil magnetic field becomes entwined forming magnetic towers with their axis coinciding with the rotational one.
Therefore, further helping the jet stay focused to drill its way out of the star.

In \citep{2002ApJ...575..996L} a binary model is proposed where angular momentum is injected back into the stellar core at the very last hour.
This is done with the idea that little angular momentum will be lost at that stage even if large wind-mass-loss rates prevail.
Plus the orbit will still be a large reservoir of angular momentum for the spin (considering an orbit which does not become too wide from the little mass loss that may still occur).  
For binaries with Case C mass transfer (after He core burning), a common envelope phase will follow.
This allows for the orbital separation to shrink at the expense of removing the H outer shell of the star.
As the companion star aproaches the He core of the supergiant, the density increases and, eventually, the later
is again confined to its shrunken Roche lobe, thus, ending the common envelope phase.
The core of the supergiant, now, likely a Wolf-Rayet (WR) star, still fills a substantial portion of its Roche lobe nonetheless.
From the tidal-synchronization timescales estimated in \citet{1975A&A....41..329Z} and \citet{1977A&A....57..383Z}, which depend on a high power (6 to 8.5) of the ratio $(a/R)$ ($a$ being the orbital separation and $R$ the stellar radius) it can be shown \citep[see, e.g.,][]{2007Ap&SS.311..177V} that the massive star will synchronize with the orbit within its remaining lifetime before GCC.

Using this model, \citet{2002ApJ...575..996L} have correctly predicted the spin of three BHs in X-ray binaries (XBs).
These were measured and reported in \citet{2006ApJ...636L.113S} and \citet{2011MNRAS.tmp.1036S}.
Using this model, and the estimated BH spins, \citet{2007ApJ...671L..41B} and \citet{2011ApJ...727...29M}, have estimated the energies available from spin for a BH to function as a GRB central engine in the known LMXBs (low-mass XBs).
These results show that most local LMXBs have too much energy and thus are likely to blow apart the accretion disk feeding the central engine too quickly.
Thus, they are likely candidates to explain subluminous LGRBs with very energetic hypernovae, but not Cosmological GRBs.
Instead, LMC X-3, may be a more likely candidate for a Cosmological GRB as argued in \citet{2008ApJ...685.1063B}.
There exist three BHs in HMXBs with very large spins \citep{2008ApJ...679L..37L,2009ApJ...701.1076G,2011ApJ...742...85G}.  
If these massive BHs ($10 - 20 \msun$) were born with these spins the energy stored in them is several times $10^{54}$ to $10^{55}$ erg.  
These formation of these BHs would have likely disrupted the binary had they had such energy available at that time.
It is highly likely the spins of these BH were acquired after GCC, however this assumption also carries difficulties  \citep{2008ApJ...689L...9M,2011MNRAS.413..183M}.

\citet{2011arXiv1105.4193W} and \citet{2012ApJ...752...32W} suggest that magnetic torques will spin down the core of the massive star as it differentially rotates after it shrinks to start a new burning stage. 
This would prevent the star from becoming a collapsar and, thus, from being a GRB progenitor.


\section{The Collapsar Model and the Blandford-Znajek Mechanism}
\label{sec:CMandBZ}

In the Collapsar model, the Fe core of a star collapses, fails to launch a supernova (SN), and forms a BH acquiring a fraction of the angular momentum.  
The material along the rotational axis falls directly in a dynamical timescale, clearing a path for the GRB.  
The material which is centrifugally supported forms an accretion disk.

A strong magnetic field ($B > 10^{14}$ G) permeates the plasma around the BH.  
The BH spins rapidly, pulling along the magnetic field.
The infalling plasma, forming the accretion disk, cannot keep up with the rotation of the BH and the Blandford-Znajek (BZ) mechanism sets in.
The Poynting flux is directed towards the rotational axis, but, the average Poynting flux is parallel to it, thus,
a jet is launched in the rotational axis direction, which, as mentioned above, has lower density.
Eventually, the BZ mechanism deposits so much energy in the disk that it explodes as a hypernova (HN).

\section{Synchronization Timescales}
\label{sec:Synch}

\begin{table}
\begin{center}
\begin{tabular}{|c|c|c|c|}
\hline
      Layer     &       Mass      &      Radius      &           Density        \\
   Composition  &   $ [\msun]$    &    $[\rsun]$     &    $[\rm g \;cm^{-3}]$   \\
\hline
\hline
        He      &        $5$      &      $10^{11}$   &   $\frac{3}{4\pi}10^4$   \\
       C-O      &       $10$      &      $10^{10}$   &   $\frac{6}{4\pi}10^7$   \\
        Fe      &      $1.5$      &       $10^8$     &   $\frac{9}{4\pi}10^9$   \\
\hline
\end{tabular}
\end{center}
\caption{Toy model of a pre-core-collapse helium star of $16.5 \msun$.  Loosely based on the mass ratios in Fig. 33.1 of \citet{1990sse..book.....K}}\label{Tab:ModelStar}  
\end{table}

Tidal-synchronization:
From \citet{1975A&A....41..329Z} and  \citet{1977A&A....57..383Z} we know that the dynamical-tide-induced-synchronization timescale and equilibrium-tide synchronization timescales are
\be
\tau_{DT} \propto \left(\frac{a}{R}\right)^{17/2}\;\;\;\;\;{\rm and}\;\;\;\;\;\tau_{ET} \propto \left(\frac{a}{R}\right)^6,
\ee
respectively, where $a$ is the orbital separation and $R$ is the stellar radius.

For the Case-C-mass-transfer binary progenitors of GRBs we propose here, these synchronization timescales translate to a few thousand years \citep[see, e.g.,][]{2007Ap&SS.311..177V}, or about the time left before GCC.

Alfv\'en:
From Jackson (1967, Classical Electrodynamics) we know the Alfvén timescale is:
\be
\tau_{\rm A}\!=\!\frac{R_c\sqrt{4\pi\rho}}{B},
\ee
here $R_c$ is the radius of the core, $\rho$ is the density and $B$ is the magnetic field that threads the core.

\begin{table}
\begin{center}
\begin{tabular}{|c|c|c|c|}
\hline
     Final B Field     &   $\tau_{\rm A,He}$   &   $\tau_{\rm A,C}$    &  $\tau_{\rm A,Fe}$  \\
    $ [\rm Gauss]$     &   $ [\rm Seconds]$    &   $[\rm Seconds]$     &   $[\rm Seconds]$   \\
\hline
\hline
        $10^{15}$      &   $5.5 \times 10^4$   &   $2.5 \times 10^3$   &       $1$        \\
        $10^{12}$      &   $5.5 \times 10^7$   &   $2.5 \times 10^6$   &       $10^3$        \\
        $10^{10}$      &   $5.5 \times 10^9$   &   $2.5 \times 10^8$   &       $10^5$        \\
\hline
\end{tabular}
\end{center}
\caption{Alfv\'en timescales for the helium, carbon and iron shells estimated for fossil magnetic fields threading the star such that the magnetic field around the compact object becomes that of the first column after core collapse.}\label{Tab:AlfvenTS}
\end{table}
\section{Generating a Magnetar strength field during Core Collapse}
\label{sec:Bfld}

From 
\be
E_B = uV = \left(\frac{B^2}{8\pi}\right)\left(\frac{4\pi R_c^3}{3}\right),
\ee
($E_B$ is the energy, $u$ is the energy density and $V$ is the collapsed-core volume of radius $R_c$ where the BZ mechanism will be active; see Moreno M\'endez 2013, ApJ submitted) one can easily estimate that the energy required
to produce a magnetar-strength magnetic field is between 10$^{-4}$ and 10$^{-7}$ times the energy available during the GCC and the end of the GRB/HN phase.  
Thus it is not unlikely that the BZ-required magnetic fields may be generated at this late stage.

\section{Conclusions}
\label{sec:Concl}

It has been shown that a massive star in a binary that undergoes Case-C mass transfer, a common envelope and tidal synchronization may retain most of the core angular momentum gained through tidal interactions if the magnetic field threading the stellar core and the He envelope is around 100 G.  Furthermore, a magnetar-like field, 10$^{15}$ G, may be easily produced during core collapse (e.g., via a SASI or convective dynamos) given the large amount of available energy during this phase of stellar evolution ($\sim10^{53}$ to $10^{54}$ erg, as opposed to $10^{47}$ to $10^{50}$ erg to generate the field).
This scenario allows a BZ engine to power GRBs and is capable of explaining Cosmological as well as subluminous GRBs.  This model also presents 7 known Galactic XRBs as relics of subluminous GRBs as well as LMC X-3 as a relic of a Cosmological GRB.


\section*{Acknowledgements}
\label{sec-Acknow}

EMM had support from a CONACyT fellowship and projects CB-2007/83254 and CB-2008/101958.  
This research has made use of NASA’s Astrophysics Data System as well as arXiv.


\end{document}